\documentclass{Interspeech2024}
\usepackage{multirow}

\interspeechcameraready

\title{TCG CREST System Description for the Second DISPLACE Challenge }

\name[affiliation={1,3}]{Nikhil}{Raghav}
\name[affiliation={1,2}]{Subhajit}{Saha}
\name[affiliation={1,2}]{Md}{Sahidullah}
\name[affiliation={1,2,4}]{Swagatam}{Das}

\address{
  $^1$Institute for Advancing Intelligence, TCG CREST, Kolkata-700 091, India\\
  $^2$Academy of Scientific and Innovative Research (AcSIR), Ghaziabad-201 002, India \\
  $^3$Department of Computer Science, RKMVERI, Howrah-711 202, India \\ $^4$ Electronics and Communication Sciences Unit, Indian Statistical Institute, Kolkata-700 108, India}
\email{nikhil.raghav.92@tcgcrest.org}


\begin{document}

\maketitle
\begin{abstract} 
In this report, we describe the speaker diarization (SD) and language diarization (LD) systems developed by our team for the Second DISPLACE Challenge, 2024. Our contributions were dedicated to Track 1 for SD and Track 2 for LD in multilingual and multi-speaker scenarios. We investigated different speech enhancement techniques, voice activity detection (VAD) techniques, unsupervised domain categorization, and neural embedding extraction architectures. We also exploited the fusion of various embedding extraction models. We implemented our system with the open-source SpeechBrain toolkit. Our final submissions use spectral clustering for both the speaker and language diarization. We achieve about $7\%$ relative improvement over the challenge baseline in Track 1. We did not obtain improvement over the challenge baseline in Track 2.
\end{abstract}

\section{Notable highlights}
Speaker diarization (SD) is the task of determining ``who spoke when" in a multi-speaker audio recording \cite{anguera2012speaker, park2022review}. On the other hand, language diarization is the task of answering ``which language was spoken when'' in multi-lingual conversations \cite{6625316}. 
The \textit{DIarization of SPeaker and LAnguage in Conversational Environments} (DISPLACE) challenge 2024 focuses on exploiting these models in a multi-lingual setting. To identify the performance of these systems in scenarios of code-mixing and code-switching \cite{unknown}. The challenge includes three tracks dedicated to SD in multilingual scenarios (Track 1), LD in multi-speaker settings (Track 2) and automatic speech recognition (ASR) in multi-accent settings (Track 3). Our team participated in Track 1 and 2. 

In this challenge, we explored different components of speaker diarization starting from speech enhancement to re-segmentation. We explored different statistical speech enhancement algorithms~\cite{loizou2007speech}. For speech activity detection (SAD), we evaluated statistical SAD such as rVAD~\cite{tan2020rvad} as well as neural SAD such as Pyannote~\footnote{\url{https://github.com/pyannote/pyannote-audio}} and Silero\footnote{\url{https://github.com/snakers4/silero-vad}}. For embedding extraction, we experimented with different embedding extractors trained with supervised as well as self-supervised manner. Similarly for clustering~\cite{han2011data}, we explored agglomerative hierarchical clustering (AHC) and spectral clustering (SC). Finally, for re-segmentation, we employed variational Bayes - hidden Markov model (VB-HMM)~\cite{landini2022bayesian}. We conducted experiments with SpeechBrain toolkit~\cite{speechbrain}. We have modified the recipe provided for AMI corpus~\footnote{\url{https://github.com/speechbrain/speechbrain/tree/develop/recipes/AMI}} and implemented for the evaluation of DISPLACE challenge dataset. For re-segmentation, we adopted the Kaldi-based scripts available with the baseline system for the third DIHARD challenge~\footnote{\url{https://github.com/dihardchallenge/dihard3_baseline}}.

Our final submission in Track 1 consists of ECAPA-TDNN embeddings extracted with a pre-trained model trained on the VoxCeleb corpus. On the other hand, for Track 2, we fused the affinity scores obtained with embeddings from two types of embedding extractors: one trained in a supervised way with ECAPA-TDNN architecture, and another trained in a self-supervised manner with XLS-R architecture.

\section{Data resources}
We have used SpeechBrain toolkit for implementing the primary speaker diarization system. We also use the baseline provided with the third DIHARD challenge for implementing the VB-HMM based re-segmentation method. In Table~\ref{Track1Config} and~\ref{Track2Config2}, we summarize the architectures of the speaker diarization components and corresponding data resources.

\begin{table}[h]
\centering
\begin{tabular}{|c|c|c|}
\hline
Component & Architecture & Speech data \\
\hline
Speech activity detector  & Pyannote & DIHARD3 \\
\hline
Embedding extractor & ECAPA-TDNN & VoxCelebs \\
\hline
Re-segmentation & VB-HMM &  VoxCelebs\\
\hline
\end{tabular}
\caption{Speaker diarization components used in \textbf{Track 1} and corresponding training data.}
\label{Track1Config}
\end{table}

\begin{table}[h]
    \centering
    \begin{tabular}{|c|c|c|}
        \hline
        Component & Architecture & Speech data \\
        \hline
        Speech activity detection  & Pyannote & DIHARD3 \\
        \hline
        \multirow{1}{*}{Embedding extractor I} & \multirow{1}{*}{ECAPA-TDNN} & \multirow{1}{*}{VoxLingua107} \\
        \hline
        \multirow{6}{*}{Embedding extractor II} & \multirow{7}{*}{XLS-R} & VoxPopuli \\
        & & MLS \\
        & & CommonVoice \\
        & & VoxLingua107 \\
        & & Babel \\
        \hline
    \end{tabular}
    \caption{Language diarization components used in \textbf{Track 2} and corresponding training data.}
    \label{Track2Config2}
\end{table}


\section{Detailed description of the algorithm}
We have used fixed segmentation of maximum length 2.0s and overlap 0.4s. Track 1 submission uses a single system as described in Table~\ref{Track1Config}. The Track 2 submission as described in Table~\ref{Track2Config2} uses affinity matrix fusion with weight 0.8 for affinity matrix from ECAPA-TDNN embedding and 0.2 for affinity matrix from XLS-R embedding. Note that to represent segment-level embeddings, we have performed average pooling on the frame-level embeddings computed with XLS-R.

\section{Results on the development and evaluation set}

Table~\ref{tab:diarization_performance} summarizes the results of the submitted system for both tracks. We have also compared with the baseline system developed by the challenge organizers. Our submissions in both tracks show competitive diarization error rates (DERs).

\begin{table}[h]
    \centering
    \begin{tabular}{|c|c|c|c|c|}
        \hline
        & \multicolumn{2}{c|}{Baseline} & \multicolumn{2}{c|}{Proposed systems}\\
        \hline Track 
        & Dev & Eval & Dev & Eval\\
        \hline
        1 & \textbf{29.16} & 34.76 & 29.61 & \textbf{32.22} \\
        \hline
        2 & 40.58 & \textbf{32.68} & \textbf{38.16} & 43.76 \\
        \hline
    \end{tabular}
    \caption{Performance of the submitted systems and the baseline systems for Track 1 and Track 2 in terms of the diarization error rate.}
    \label{tab:diarization_performance}
\end{table}


\section{Hardware requirements}
We conduct the experiments using a machine consisting CPUs with 12 cores and 24 threads, equipped with an AMD Ryzen 9 7900X (24) @ 4.7GHz. The GPU utilized was the NVIDIA GeForce RTX 4090. The system includes 128 GB of RAM and ran on the Ubuntu 22.04.3 operating system. 


\section{Conclusion}
In this challenge, we explored different components of a speaker diarization pipeline. Our initial experiments with different speech enhancement methods did not significantly improve the performance over systems without speech enhancement techniques. We also attempted to extract SAD labels from the enhanced speech, but it did not help. Due to resource and time constraints, we used publicly available pre-trained models only for neural speech activity detection and embedding extraction. We also used the default parameters for clustering and re-segmentation. Our future plan is to investigate fine-tuning of the pre-trained model and further parameter optimization.

\bibliographystyle{IEEEtran}
\bibliography{mybib}

\end{document}